\documentclass[11pt]{article}
\usepackage{blois,epsfig}
\usepackage[latin1]{inputenc}
\usepackage[OT1]{fontenc}

\def\be{\begin{equation}}
\def\ee{\end{equation}}
\def\bea{\begin{eqnarray}}
\def\eea{\end{eqnarray}}

\begin{document}
\vspace*{4cm}
\title{Toward directional detection of Dark Matter with the DM-TPC detector} 

\author{ G.~SCIOLLA$^{1,~}$\footnote{Corresponding author: sciolla@mit.edu}, 
S.~AHLEN$^2$, 
D.~DUJMIC$^1$, 
V.~DUTTA$^1$, 
P.~FISHER$^1$, 
S.~HENDERSON$^1$, 
A.~KABOTH$^1$, 
G.~KOHSE$^1$, 
R.~LANZA$^1$, 
J.~MONROE$^1$, 
A.~ROCCARO$^2$, 
N.~SKVORODNEV$^3$, 
H.~TOMITA$^2$, 
R.~VANDERSPEK$^1$, 
H.~WELLENSTEIN$^3$, 
R.~YAMAMOTO$^1$\\ 
(The DM-TPC Collaboration) }
\address{
$^1$ Massachusetts Institute of Technology, Cambridge, MA 02139 (USA)\\
$^2$ Boston University, Boston, MA 02215 (USA) \\
$^3$ Brandeis University,  Waltham, MA 02454 (USA)
}
\maketitle\abstracts{
Directional detection can provide unambiguous observation of 
Dark Matter interactions even in 
presence of insidious backgrounds.
The DM-TPC collaboration is developing a detector with the goal 
of measuring the direction and sense 
of nuclear recoils produced in  Dark Matter interactions. 
The detector consists of a  Time Projection Chamber with optical 
readout filled with CF$_4$ gas at low pressure. A collision between a WIMP and 
a gas molecule results in a nuclear recoil of 1-2 mm. 
The measurement of the energy loss along the recoil allows us to 
determine the sense and the direction of the recoil.
Results from a prototype detector operated in a low-energy neutron beam
clearly demonstrate the suitability of this approach to measure
directionality. A cubic meter prototype, which is now being designed, will
 allow us to set competitive limits on
spin-dependent Dark Matter interactions using a directional detector.
}

\section{The need for a novel Dark Matter detector}

Searches for non-baryonic Dark Matter (DM) in the form of Weakly Interacting Massive Particles (WIMPs) rely on detection of nuclear recoils created by the elastic scattering between a WIMP and the detector material. 
In presence of backgrounds, an unambiguous positive observation can be provided by 
detecting the direction of the incoming WIMP.\cite{directionality}  
As the Earth moves in the galactic 
Dark Matter halo, the WIMPs appear to come at us with an average velocity of 220 km/s from 
the direction of the constellation Cygnus. Due to the relative orientation between the 
Earth's rotation axis and the direction of the Dark Matter wind, 
the direction of the WIMPs changes with respect to our detector on Earth by about 90 degrees every twelve hours. 
Since no background is expected to correlate with the position of Cygnus in the sky, 
directional detection will  improve  the sensitivity to Dark Matter by orders of magnitude.\cite{agreen}
If the detector is also able to determine the sense of the direction, 
the sensitivity to DM is further enhanced.\cite{agreen} 
In addition to background rejection, the measurement of the direction of Dark Matter 
will also allow us to discriminate between various Dark Matter models. 

Dark Matter particles can interact with ordinary matter via spin-independent (scalar) or spin-dependent (axial vector) interactions. 
 Most of the current experiments concentrate on searches for scalar couplings and are able to place very stringent limits on spin-independent interactions excluding cross-sections above $10^{-43}$ cm$^2$.  
In contrast, the existing limits on axial-vector couplings are about seven orders of magnitude less stringent. 
Despite the modest experimental effort in this sector, these interactions are interesting because they are expected 
to be enhanced with respect to the scalar interactions in theoretical models in which the Lightest Supersymmetric Particle has a substantial Higgsino contribution.\cite{ref5} Therefore there is an urgent need for improving the searches for spin-dependent interactions of Dark Matter.
Materials rich in fluorine, with nuclear spin of 1/2, are the most suitable detector materials for such searches.\cite{ref4}

\section{The DM-TPC detector }
The DM-TPC detector~\cite{ref1,ref2} consists of a low pressure Time Projection Chamber (TPC) filled with 
tetra-fluoro-methane (CF$_4$). For a pressure inside the vessel of 50 torr, the typical collision of a WIMP 
with a gas molecule causes a nucleus to recoil by about 1 mm. 
The ionization electrons produced by the recoiling nucleus drift in the gas along a uniform electric field toward an amplification region created by two parallel woven meshes.\cite{ref2} The large electric fields present in this region cause the avalanche process, during which a substantial amount of scintillation light is produced.\cite{ref3} A CCD camera is used to detect such photons and to image the projection of the nucleus parallel to the amplification plane. The total amount of light deposited in the CCD measures the total energy of the recoil. Because the energy loss is not uniform along the trajectory, we can determine not only the direction of the incoming WIMP, 
but also its sense (``head-tail'' measurement). 
A phototube provides a measurement of the recoil parallel to the drift direction and serves as a trigger for reading out the CCD camera.

The combination of the various measurements provided by this detector is very effective in suppressing backgrounds due to alpha particles, electrons, and photons. As an example, the rejection factor measured for photons is better than 2 parts per million. 

The DM-TPC detector is designed with the goal of maturing into a large underground experiment.
The choice of an optical readout is motivated by the modest cost-per-channel obtainable with the use of a CCD camera, 
making it possible in the future to economically scale the detector to large volumes. 
The choice of CF$_4$ gas as active material is motivated by the low transverse diffusion and good scintillation properties of this gas. 
In addition, CF$_4$ is non-toxic and non-flammable, making it easier to operate the detector underground. 
Finally, this gas contains four atoms of fluorine for each atom of carbon, making the DM-TPC detector ideal to study  
spin-dependent interactions. 

\section{Recent results }
The first prototype of the DM-TPC detector utilizing  an amplification region made of parallel wires
started taking data in Spring 2007 at a pressure of 200 torr. 
This small chamber  was built to demonstrate the validity of the detector technology and 
to  prove that we could indeed determine the sense of the direction of low-energy nuclear 
recoils, i.e. observe the so-called ``head-tail'' effect. 

The prototype was calibrated with 5.5 MeV alpha particles produced by a $^{241}$Am source. 
The energy loss (dE/dx) for 5.5 MeV alpha particles in CF$_4$ was measured and compared with the SRIM~\cite{SRIM} MC simulation. The agreement between data and MC was found to be excellent. The same alpha source was also used to measure the resolution as a function of the drift distance of the primary electrons to study the effect of the diffusion. These studies showed~\cite{ref1} that the drift distance should be limited to 25 cm.  

The initial measurements used  a 14 MeV neutron source from a deuteron-triton tube, and the reconstructed recoils had an energy between  200 and 800 MeV. 
Because of the small energy of the recoiling nucleus, we expect the energy deposition to uniformly decrease along the path of the recoil, allowing us to identify  the ``head'' (``tail'') of the event by a smaller (larger) energy deposition.
Such an effect was clearly observed in the data, which allowed us to publish an 8 $\sigma$  
observation of the ``head-tail'' effect in  low-energy neutrons.\cite{ref1}

\begin{figure}
{\psfig{figure=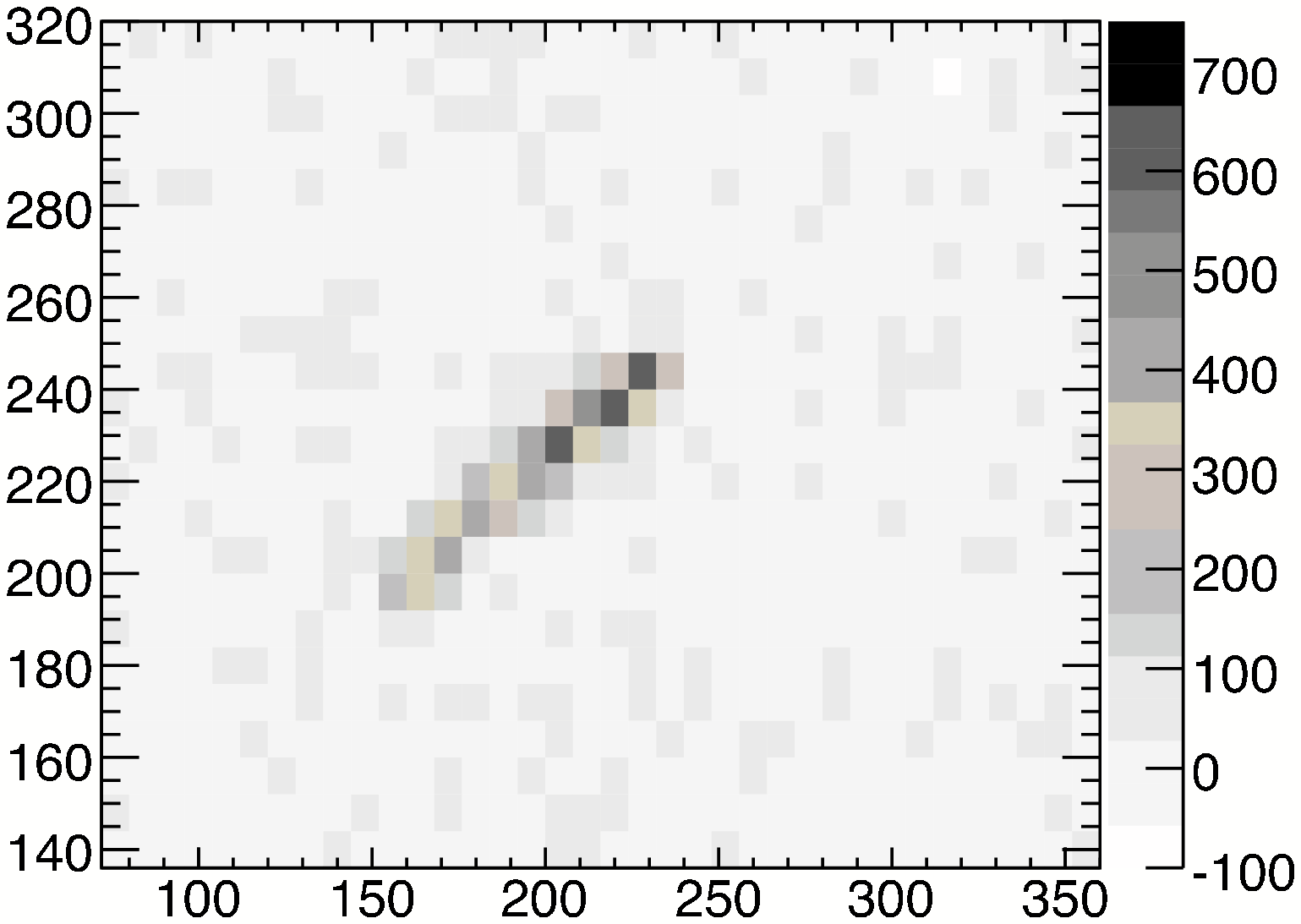,width=3.2in}}
{\psfig{figure=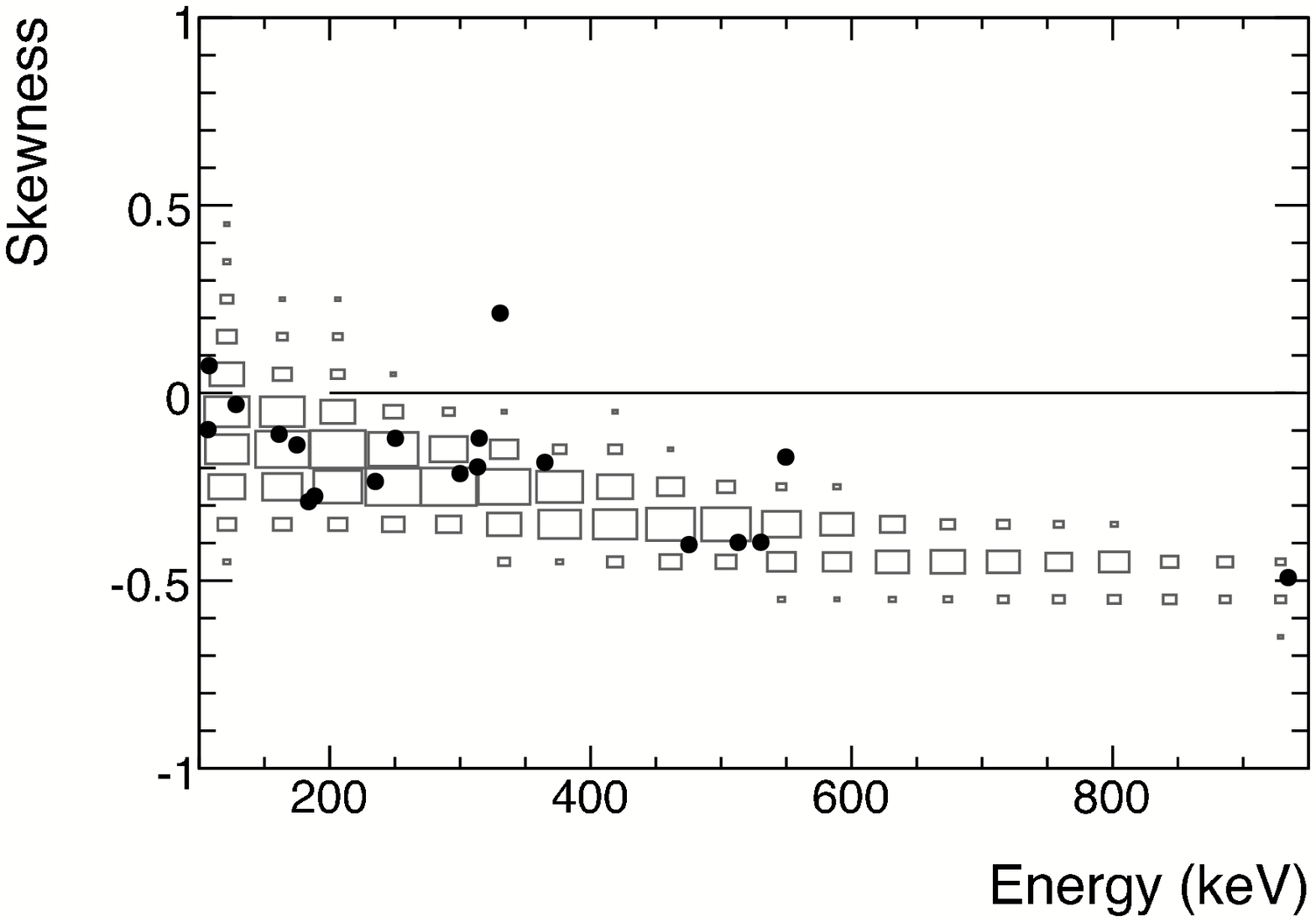,width=3.0in}}
\caption{
Left: dE/dx distribution for a fluorine nucleus recoiling with an energy of about 250 keV in CF4 at 75 torr. 
The direction of the incident neutron is right to left. The larger energy loss at the right of the track and lower 
energy loss at the left of the track is characteristics of the ``head-tail'' effect. 
Right: skewness of recoil events versus the recoil's kinetic energy. The black dots indicate data, while the 
histogram shows the Monte Carlo. 
\label{fig2}}
\end{figure}

These measurements were repeated using an improved detector employing a mesh-based amplification region.
Nuclear recoils from a $^{252}$Cf  source at a pressure of 75 torr 
proved~\cite{ref2}  that our detector has good head-tail discrimination for recoils as low as 100 keV. 
Figure \ref{fig2}(left) shows the energy-loss per unit length (dE/dx) of a typical nuclear recoil reconstructed 
in our detector. The change of dE/dx along the recoil track is clearly visible. 
To quantify the ``head-tail'' effect we define the skewness as 
$\gamma\equiv\mu_3/\mu_2^{1.5}$, where $\mu_2$ and $\mu_3$ are the second and third moments of the energy deposit 
along the track, respectively. 
Figure \ref{fig2}(right) shows skewness as a function of the energy of the recoil track. 
The black dots represent data, while the histogram shows the prediction of a Monte Carlo simulation based on SRIM~\cite{SRIM}. 
As expected,  the measurement of the skewness is better for  higher recoil energies.  

\begin{figure}
{\psfig{figure=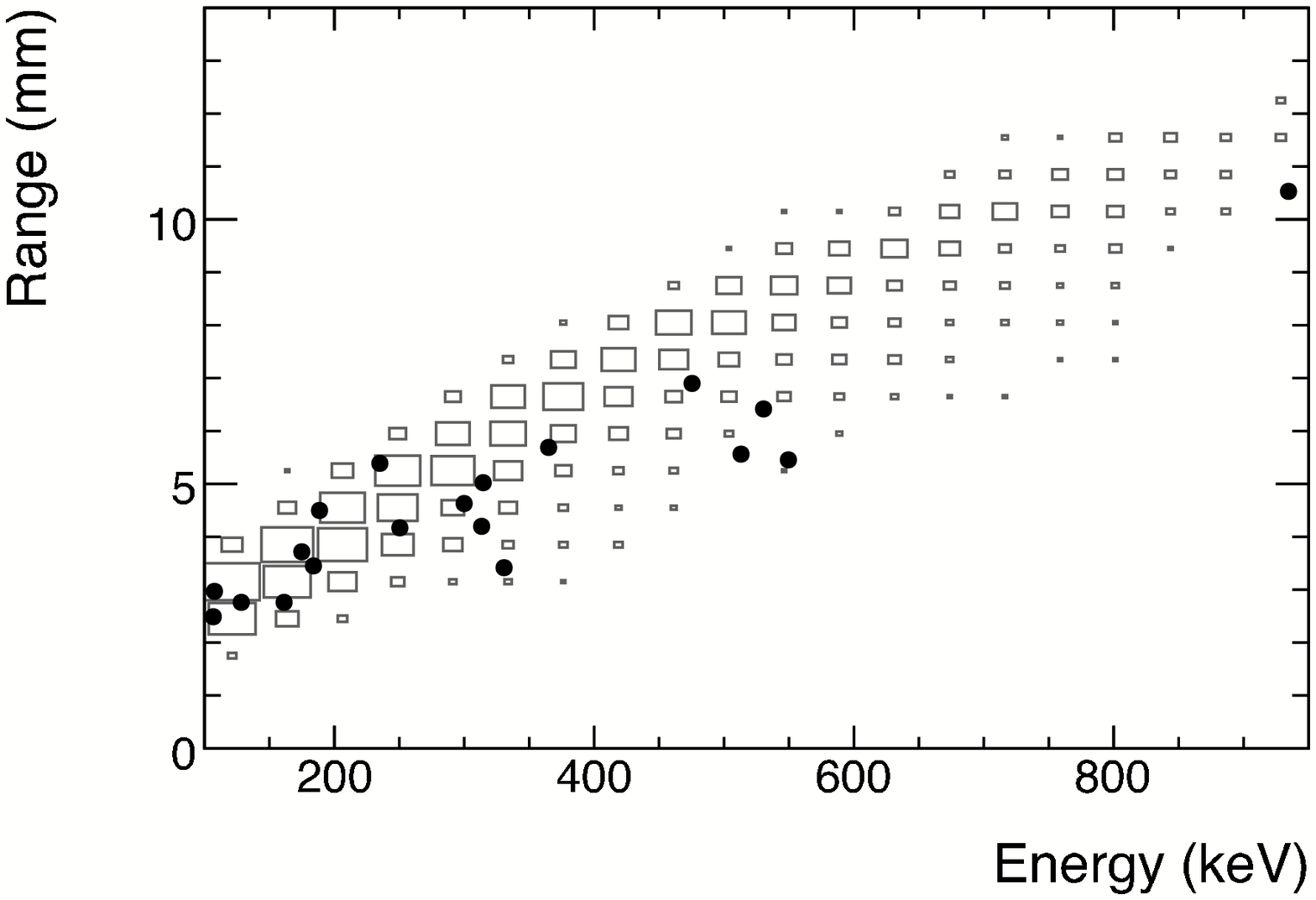,width=3.1in}}
{\psfig{figure=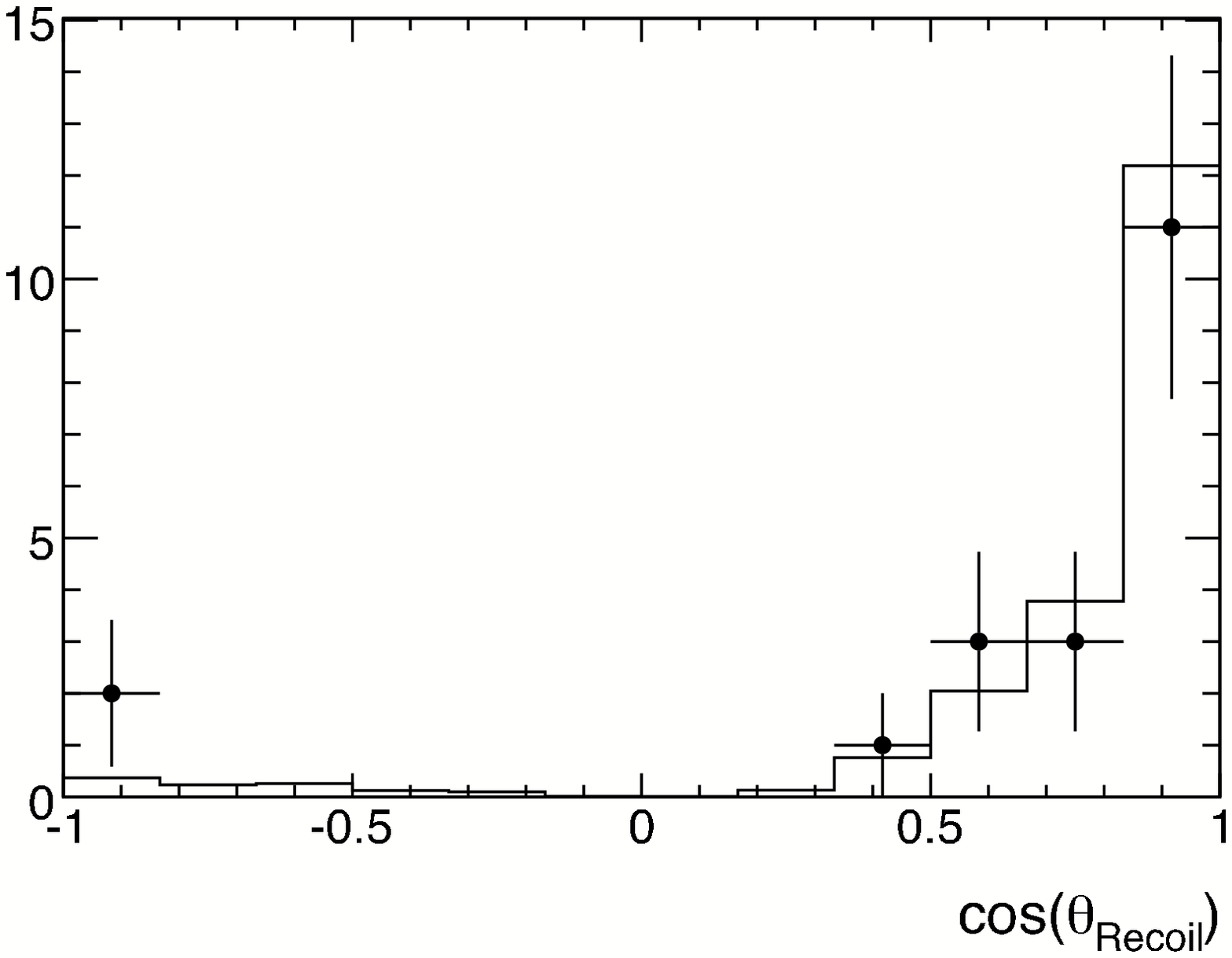,width=3.1in}}
\caption{
Left: Range  vs. reconstructed energy for nuclear recoil candidates in a $^{252}$Cf exposure at 75~Torr. 
Black points are data, the box-histogram is simulation.
Right:Signed distribution of the cosine of the 2D recoil angle, with data and simulation normalized to the same area.
\label{fig3}}
\end{figure}

Figure \ref{fig3}(left) shows the correlation between the range of the reconstructed track (y axis) and the 
energy of the recoil (x axis). The cosine of the  angle between the direction of the 
incident neutron and the direction of the nuclear recoil is shown in figure  \ref{fig3}(right). 
The agreement between data (back dots) and simulation (histogram) is excellent. 
The angular resolution for the nuclear recoils was measured to be better than 15 degrees.

\section{Underground run with a larger  detector }
The promising results obtained by our R\&D efforts encourage us to build a 
larger detector with an active volume of about one cubic meter. 
When operated at 50 torr, this device will have an active mass of 250 g. 
In one year of operation, such a detector will accumulate 90 kg-day of exposure. 
The underground operation of this detector will provide very competitive spin-dependent limits 
on Dark Matter cross-sections ($\sigma_{SD}\approx 10^{-37} cm^2$).  

If successful this effort  will lay the foundation for a large (a few hundred kg) directional Dark Matter 
detector with substantial potential to directly observe Dark Matter and determine its direction. 
This experiment is an ideal candidate for the Deep Underground Science and Engineering Laboratory (DUSEL) 
that is being planned in the Homestake mine in South Dakota.

\section{Summary and conclusion}

Directional detection may hold the key to the unambiguous observation of Dark Matter in presence of backgrounds, and allows us to discriminate between models that predict Dark Matter to come from different directions in our galaxy. 

The DM-TPC collaboration is developing a detector to achieve this goal. 
The device consists of a low-pressure TPC filled with CF$_4$ gas 
read out by an array of CCD cameras. Our prototypes proved the detector concept and demonstrated its 
ability to reconstruct both the sense and direction of nuclear recoils above 100 keV. A larger detector is being designed for underground operations in 2009 with the goal of obtaining competitive results on spin-dependent interactions using directional information. The success of this device will lay the foundation for a large Dark Matter experiment that will be able to detect the direction of WIMPs and discriminate between DM models in our galaxy.

\section*{Acknowledgments}
This work was supported by the 
Advanced Detector Research Program of the U.S. Department of Energy (contract number 6916448), the National Science Foundation, 
the Reed Award Program, the Ferry Fund, the Pappalardo Fellowship program, the MIT Kavli Institute for Astrophysics and 
Space Research, and the Physics Department at the Massachusetts Institute of Technology.

\end{document}